\documentclass[3p,times,twocolumn]{elsarticle}

\usepackage{ecrc}

\volume{00}

\firstpage{1}

\journalname{Nuclear and Particle Physics Proceedings}

\runauth{B. Duclou\'e, T. Lappi, Y. Zhu}

\jid{nppp}

\jnltitlelogo{Nuclear and Particle Physics Proceedings}

\usepackage{amssymb}
 \usepackage{amsthm}
 \usepackage{amsmath}

\usepackage[figuresright]{rotating}

\usepackage[utf8]{inputenc}

\newcommand{\rt}{{\mathbf{r}}}
\newcommand{\xt}{{\mathbf{x}}}
\newcommand{\bt}{{\mathbf{b}}}

\newcommand{\yt}{{\mathbf{y}}}

\newcommand{\pt}{{\mathbf{p}}}
\newcommand{\qt}{{\mathbf{q}}}
\newcommand{\kt}{{\mathbf{k}}}

\newcommand{\ptt}{p_\perp} 
\newcommand{\ktt}{k_\perp} 
\newcommand{\qtt}{q_\perp} 
\newcommand{\ltt}{l_\perp} 
\newcommand{\lt}{\mathbf{l}}

\newcommand{\xif}{{\xi}_\text{f}}
\newcommand{\xf}{{x}_\text{f}}

\newcommand{\ud}{\mathrm{d}}

\newcommand{\tr}{\, \mathrm{Tr} \, }

\newcommand{\nc}{{N_\mathrm{c}}}

\newcommand{\cf}{C_\mathrm{F}}

\newcommand{\qs}{Q_\mathrm{s}}

\newcommand{\as}{\alpha_{\mathrm{s}}}

\newcommand{\fig}{Fig.~}

\newcommand{\eq}{Eq.~}

\newcommand{\re}{Ref.~}

\newcommand{\ical}{\mathcal{I}}
\newcommand{\jcal}{\mathcal{J}}
\newcommand{\scal}{\mathcal{S}}

\allowdisplaybreaks

\begin{document}

\begin{frontmatter}



\dochead{}

\title{Forward hadron production in pA collisions beyond leading order }


\author{B. Duclou\'e, T. Lappi, Y. Zhu}

\address{Department of Physics, P.O. Box 35, 40014 University of Jyv\"askyl\"a, Finland \\
Helsinki Institute of Physics, P.O. Box 64, 00014 University of Helsinki,
Finland}

\begin{abstract}
In this talk, we report our recent progress on pinning down the cause of negativity in  the NLO single inclusive hadron production in pA collisions at forward rapidity.
\end{abstract}


\end{frontmatter}


\section{Introduction}
\label{intro}
Parton densities in hadrons are strongly enhanced when the hadrons are accelerated at high energy colliders. However, due to the merging of soft partons, the parton density stops increasing when the energy is large enough to reach the saturation regime. Parton density distributions at  high energy are well described by the color glass condensate (CGC) effective theory \cite{Gelis:2010nm}. To study saturation effects, the ideal way is to study processes where a dilute projectile interacts with a dense target. This motivates the intensive study of forward hadron production in proton-nucleus (pA) scatterings at high energies.

At high energy, the cross section for this process can be factorized into the convolution of a perturbatively-calculable hard part, non-perturbative coefficients describing the parton densities in the incoming hadrons and the hadronization of final-state partons into hadrons. The unintegrated gluon distribution in the dense target is related to the dipole correlator $S(\rt=\xt-\yt)=\left< \frac{1}{\nc}\tr U(\xt)U^\dag(\yt) \right>$ in momentum space. Its rapidity evolution is described by the well-known Balitsky-Kovchegov (BK) equation \cite{Balitsky:1995ub,Kovchegov:1999yj}. The leading-order (LO) cross section \cite{Dumitru:2002qt} for single inclusive forward hadron production in pA collisions has been derived, and numerical implementations \cite{Dumitru:2005gt,Albacete:2010bs,Tribedy:2011aa,Rezaeian:2012ye,Lappi:2013zma} of the LO cross section are found to be consistent with experimental measurements, however with a rather large overall normalization factor.  It is therefore instructive to understand how these results would change at higher orders.

The cross section for forward hadron production in pA collisions was calculated at next-to-leading order (NLO)  \cite{Chirilli:2011km,Chirilli:2012jd} and the first numerical implementation of these expressions showed that the cross section for the production of hadrons at large transverse momenta \cite{Stasto:2013cha} is negative. Several proposals \cite{Kang:2014lha, Altinoluk:2014eka, Watanabe:2015tja} have suggested to implement the explicit kinematic constraint or `Ioffe time' cutoff to solve this issue, which however, did not remove the unphysical negativity completely. 

In this proceeding, we analyse the cause of the negativity at large transverse momentum at NLO, and try to fix this issue. For the sake of  simplicity, we will only consider the $q\rightarrow q$ channel in the following as it has a similar behavior as the full NLO cross section.  We will use the simple Golec-Biernat and W\"usthoff (GBW) \cite{GolecBiernat:1998js} parametrization for the dipole correlator, and the correction from NLO BK equation will be discussed at the very end.

\section{Formalism}
\label{form}
The NLO cross section  for single inclusive hadron production at forward rapidity can be read from \re\cite{Chirilli:2012jd} which will be referred to as `CXY'. By leaving out an overall integration over the impact parameter $\bt$, the differential multiplicity for the quark channel reads
\begin{eqnarray}\label{eq:nlosigma}
&& \hspace{-1.4cm}\frac{\ud N^{pA\to hX}}{\ud^2\pt \ud y_h}
=
\int_\tau^1 \frac{\ud z}{z^2}D_{h/q}(z) x_p q(x_p) \frac{\scal^{(0)}(\ktt) }{(2\pi)^2}
\\ \nonumber
&& +  \frac{\as}{2\pi^2} \int \frac{\ud z}{z^2}D_{h/q}(z) 
\int_{\tau/z}^1 \ud \xi \frac{1+\xi^2}{1-\xi}
\frac{x_p}{\xi} q\left(\frac{x_p}{\xi}\right) \\ \nonumber
&&\times \left\{\cf \ical(\ktt,\xi) + \frac{\nc}{2}\jcal(\ktt,\xi) \right\}
\\ \nonumber
&&- \frac{\as}{2\pi^2} \int \frac{\ud z}{z^2}D_{h/q}(z) 
\int_{0}^1 \ud \xi \frac{1+\xi^2}{1-\xi}
x_p q\left(x_p \right) \\ \nonumber
&&\times\left\{\cf \ical_v(\ktt,\xi) + \frac{\nc}{2}\jcal_v(\ktt,\xi) \right\},
\end{eqnarray}
where 
\begin{eqnarray}
&&\hspace{-1.4cm}\ical(\ktt,\xi) =
\int \frac{\ud^2 \qt}{(2\pi)^2} \scal(\qtt)
\left[\frac{\kt-\qt}{(\kt-\qt)^2} - \frac{\kt-\xi \qt}{(\kt-\xi \qt)^2} \right]^2\;,
\\ 
&&\hspace{-1.4cm}\jcal(\ktt,\xi) =
\int \frac{\ud^2 \qt}{(2\pi)^2} \scal(\qtt) \bigg[
\frac{2(\kt-\xi\qt)\cdot(\kt-\qt)}{(\kt-\xi\qt)^2(\kt-\qt)^2}
\\ \nonumber
&&
-\int \frac{ \ud^2\lt}{(2\pi)^2}
\frac{2(\kt-\xi\qt)\cdot(\kt-\lt)}{(\kt-\xi\qt)^2(\kt-\lt)^2}
\scal(\ltt)\bigg] \;,
\\
&&\hspace{-1.4cm}\ical_v(\ktt,\xi) =
\scal(\ktt)\int \frac{ \ud^2 \qt }{(2\pi)^2}
\left[\frac{\kt-\qt}{(\kt-\qt)^2} - \frac{\xi\kt-\qt}{(\xi \kt-\qt)^2} \right]^2 , 
\\
&&\hspace{-1.4cm}\jcal_v(\ktt,\xi) =
\scal(\ktt)
\bigg[
\int \frac{\ud^2 \qt}{(2\pi)^2} 
\frac{2(\xi\kt-\qt)\cdot(\kt-\qt)}{(\xi\kt-\qt)^2(\kt-\qt)^2} \\ \nonumber
&&
-\int \frac{\ud^2 \qt}{(2\pi)^2} \frac{  \ud^2\lt}{(2\pi)^2}
\frac{2(\xi\kt-\qt)\cdot(\lt-\qt)}{(\xi\kt-\qt)^2(\lt-\qt)^2}
\scal(\ltt)
\bigg]\;
\end{eqnarray}
with
$\pt = z\kt$, $x_p = \ptt e^{y_h}/(z\sqrt{s})$, $\tau = z x_p$, $x_g = \ptt/(z\sqrt{s})e^{-y_h}$, $\ptt=|\pt|$, $\qtt=|\qt|$, $\ktt=|\kt|$, and $\ltt=|\lt|$. The color dipole in momentum space is 
$
\scal(\ktt)=\scal(\ktt,\bt)=\int \ud^2\rt e^{-i\kt\cdot\rt} S(\rt)
$, while the superscript $(0)$ stands for the tree level unrenormalized color dipole.
In these expressions, $\xi$ is the longitudinal momentum fraction of the incoming quark carried by the outgoing one after the radiation of a gluon with longitudinal momentum fraction $1-\xi$.

Apparently, there are two types of divergences in the NLO multiplicity: the collinear divergence and the rapidity divergence, occurring in terms proportional to $\cf$ and $\nc$ respectively. The collinear divergence is regulated by dimensional regularization and absorbed into the DGLAP evolution of  PDFs $q\left(x \right) $ and FFs $D_{h/q}(z)$. The rapidity divergence in the $\nc-$terms arises when $\xi\rightarrow 1$, i.e., emitting a soft gluon collinear to the target. Naturally, this divergence should be put into the evolution of the dense target. In CXY, the rapidity divergence is regulated by introducing the renormalized dipole
\begin{multline}\label{eq:cxysub}
\scal(\ktt)  =  \scal^{(0)}(\ktt)
\\ + 2 \as \nc  \int_0^1 \frac{\ud \xi}{1-\xi} 
\left[\jcal(\ktt,1) - \jcal_v(\ktt,1)\right] ,
\end{multline}
which in coordinate space is the integral form of BK equation. This procedure successfully absorbs the rapidity divergence into the evolution of the target, however it leads to the negativity of the NLO cross section at large transverse momenta. An oversubtraction in the rapidity regularization is found to be the cause of this negativity. This is easily demonstrated by the behaviour of the $\nc-$terms at large transverse momenta:  
$
 \jcal(\ktt,\xi) -
 \jcal_v(\ktt,\xi) \sim \frac{\xi}{\ktt^4}
$
, a positive and linearly increasing function of $\xi$. To reduce the oversubtraction, one can, instead of \eq(\ref{eq:cxysub}), introduce a factorization scale $\xif$ between 0 and 1 in the rapidity regularization,
\begin{multline}\label{eq:xifsub}
\scal(\ktt)  =  \scal^{(0)}(\ktt)
\\+ 2 \as \nc \int_{\xif}^1 \frac{\ud \xi}{1-\xi} 
\left[\jcal(\ktt,1) - \jcal_v(\ktt,1)\right] .
\end{multline}
This reduces to the CXY subtraction for $\xif=0$, while less positive contributions will be included in the renormalized dipole for $\xif>0$ at large $\ktt$.

\section{Results}
\label{res}
Having now the general definition of the renormalized dipole with $\xif$, we will demonstrate its effect on the cross section by varying the regularization scale. For simplicity, we will use the GBW model \cite{GolecBiernat:1998js} for the dipole correlator, which is
\begin{equation}
S(\rt)=e^{-\rt^2 \qs^2/4} \, , \quad \scal(\ktt)=\frac{4\pi}{\qs^2}e^{-\ktt^2/\qs^2} , 
\end{equation}
with the saturation scale $\qs^2=c A^{1/3} Q_{s 0}^2 \left(\frac{x_0}{x}\right)^{\lambda}$, and $A$ being the atomic number of the target nucleus, $c=0.56$, $Q_{s 0}=1$ GeV, $x_0=3.04 \times 10^{-4}$ and $\lambda=0.288$. The simplified result for the NLO cross section in the GBW model is expressed in~\cite{Ducloue:2016shw} for finite $\nc$ rather than in the large $\nc$ limit taken in CXY. This allows for a clear separation of the collinear divergence in the $\cf-$terms and the rapidity divergence in the $\nc-$terms. We use the DSS~\cite{deFlorian:2007aj} and MSTW 2008~\cite{Martin:2009iq} NLO parametrizations for the FFs $D_{h/q}(z)$ and quark PDFs $q(x)$ respectively. We will evaluate the multiplicity at RHIC energy such that $\sqrt{s}=200$ GeV, $\alpha_s=0.2$, $\mu^2=10$ GeV$^2$ and $y_h=3.2$. 

\begin{figure}[ht]
	\centering
	\includegraphics[scale=1.4]{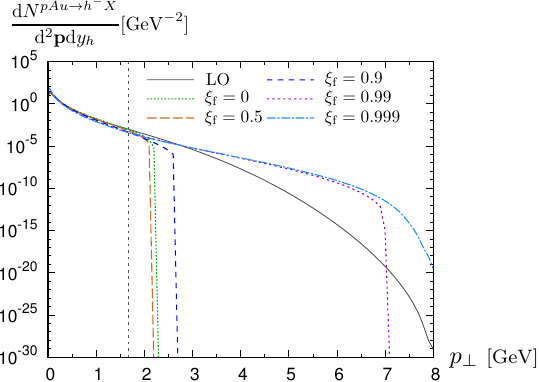}\\
	\vspace{0.6cm}
	\includegraphics[scale=1.5]{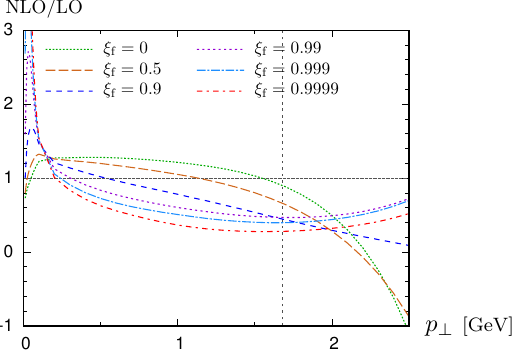}
	\caption{Upper: Multiplicity as a function of $\ptt$ for different values of $\xif$. Lower: Ratio of the multiplicity at NLO and LO for different values of $\xif$. In both cases the vertical dashed line corresponds to $\qs \approx \ptt$.}
	\label{fig:cutoff}
\end{figure}

\begin{figure}[ht]
  \centering
  \includegraphics[width=0.6\linewidth]{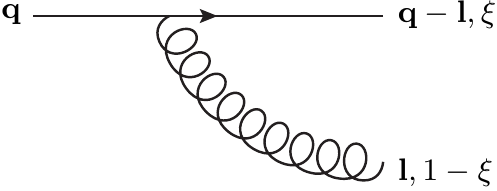}
  \caption{Gluon emission.}
  \label{fig:qqg}
\end{figure}

The results with fixed values of $\xif$ are shown in \fig\ref{fig:cutoff}. The vertical line in the figure indicates the point where $\ptt \approx Q_s(x_g)$. On the upper panel, we show the multiplicity as a function of transverse momentum for different values of $\xif$. The LO result is shown as the black solid line. For $\xif=0$, the NLO result turns negative at $\ptt\sim2.5$ GeV, which is  similar to the full NLO cross section displayed in Ref.~\cite{Stasto:2013cha}. However, as discussed in the preceding section, the oversubtraction is indeed reduced when $\xif\rightarrow 1$, in particular, the multiplicity is positive up to 8 GeV when $\xif=0.999$. In the lower panel, we show the ratio of  the NLO multiplicity to the LO one with various values of $\xif$ for $\ptt$ up to 2.5 GeV. Contrary to $\xif=0$, the NLO multiplicity is suppressed at moderate transverse momenta with respect to the LO result when $\xif$ is close to 1. 

We have seen that the value of $\xif$ indeed has an important effect on the NLO multiplicity. However the choice of $\xif$ was kind of arbitrary. Physically, the value of $\xif$ should be fixed by light cone ordering at NLO. More precisely, this is related to the $k^-$ or $x^+$ ordering required in the BK evolution. In the calculation of the NLO cross section, $k^+$ ordering is automatically implemented by the radiation of a gluon with $0<1-\xi<1$ as depicted in \fig\ref{fig:qqg}, while the $k^-$ ordering has to be implemented explicitly.  At LO, the light cone energy required from the target is
$ k^-_\text{LO} = \frac{\kt^2}{2k^+} = \frac{\kt^2}{2 x_p P^+}
$
for a collinear quark from the projectile with energy $k^+$, which defines the momentum fraction $x= k^-_\text{LO}/P^-=x_g$ from the target at LO. At NLO, according to \fig\ref{fig:qqg}, the light cone energy required for the production of the on-shell outgoing quark and gluon is
\begin{equation}\label{eq:kminusqg}
\Delta k^-_{qg} =\frac{x_g P^-}{\kt^2} \frac{(\lt-(1-\xi)\qt)^2}{\xi(1-\xi)}.
\end{equation}
A correct renormalization of the target should include the point $\xi=1$ in the subtraction. In addition, one should also implement $k^-$ ordering for the evolution of target. This can be done by requiring the light cone energy to be greater than the factorization fraction of target energy $x_f$, i.e., $\Delta k^-_{qg}>\geq \xf P^- $, which tells that gluons with $x>\xf$ should be included in the evolution of the target. A natural choice for the renormalization scale is $\xf\approx x_g$ such that all the large energy logarithms are resummed into the target.  For the production of a quark with large transverse momentum $\kt\gg Q_s$, the momentum $\qt-\lt\sim\kt\gg Q_s$. The momentum $\lt$ is an integration variable, thus of the order of $Q_s$. With those in mind, when $\xi\rightarrow 1$, the light cone ordering reduces to 
\begin{equation}
\Delta k^-_{qg} = 
\frac{x_g P^-}{\kt^2} \frac{\qs^2}{1-\xi} \geq \xf P^- \Leftrightarrow 1-\xi\leq \frac{\qs^2}{\kt^2} \frac{x_g}{\xf}. 
\end{equation}
Therefore, the factorization scale in \eq\ref{eq:xifsub} can be fixed with
\begin{equation}\label{eq:xifkt}
\xif\rightarrow\xif(\ktt)=1-\min(1,\frac{x_g}{\xf}\frac{Q_s^2}{\kt^2})\;,
\end{equation}

\begin{figure}[ht]
	\centering
	\includegraphics[scale=1.4]{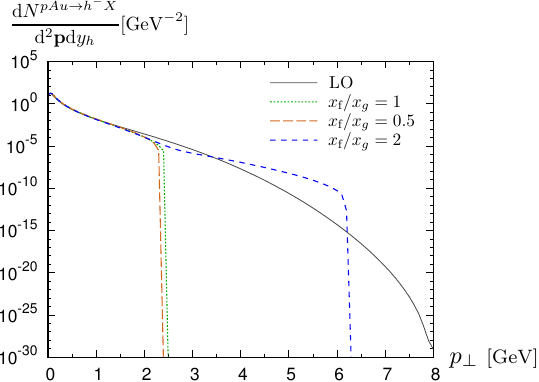}
	\caption{Multiplicity obtained using different values of $\frac{\xf}{x_g}$.}
	\label{fig:xifcontinuous}
\end{figure}

Using $\xif(\ktt) = 1/(1+ \frac{x_g}{\xf}\frac{\qs^2}{\ktt^2})$ for a smooth transition between large and small $\kt$, the result is displayed in \fig\ref{fig:xifcontinuous} with the variation of $\xf/x_g$ by a factor of 2. The choice $\xf/x_g=2$ pushes the negativity to much larger value of $\ptt$ than $\sim$2.5 GeV with CXY subtraction.

\section{Conclusions and discussion}
\label{dis}
We have studied rapidity factorization in NLO single inclusive hadron production in pA collisions at forward rapidity following the calculation of Ref.~\cite{Chirilli:2012jd}. 
The oversubtraction in the rapidity regularization is reduced by introducing a rapidity factorization scale, which allows us to achieve a positive NLO multiplicity up to arbitrarily large transverse momenta. To fix the factorization scale, we proposed to impose light cone ordering in the evolution of the target, which can significantly improve the results for larger transverse momenta. However these results are very sensitive to the variation of the factorization scale even in its natural range. This could probably be improved by using dipole correlators obeying the BK equation instead of the simple GBW model. Furthermore, light cone ordering should be imposed exactly in the transverse momentum integrals. We will address these issues in the future.

\section*{Acknowledgements} 
We thank E. Iancu, Z. Kang, B.-W Xiao and D. Zaslavsky for discussions. This work has been supported by the Academy of Finland, projects 
267321, 273464 and 303756 and by the European Research Council, grant ERC-2015-CoG-681707.

\bibliographystyle{elsarticle-num}
\bibliography{Zhu_Y}







\end{document}